\documentclass[twocolumn,showpacs,preprintnumbers,amsmath,amssymb]{revtex4}
\usepackage{graphicx}
\input epsf
\usepackage{dcolumn}
\usepackage{bm}

\begin{document}

\title{Quantum cosmologies with varying speed of light and the $\Lambda$ problem}
\author{A.V. Yurov}
\email{artyom_yurov@mail.ru}
\author{V.A. Yurov}%
 \email{yurov@freemail.ru}
 \affiliation{%
The Theoretical Physics Department, Kaliningrad State
University,A. Nevskogo str., 14, 236041,
 Russia.
\\
}
\date{\today}
\begin{abstract}
In quantum cosmology the closed universe can spontaneously
nucleate out of the state with no classical space and time. For
the universe filled with a vacuum of constant energy density the
semiclassical tunneling nucleation probability can be estimated as
$\emph{P}\sim\exp(-\alpha^2/\Lambda)$ where $\alpha$=const and
$\Lambda$ is the cosmological constant, so once it nucleates, the
universe immediately starts the de Sitter inflationary expansion.
The probability $\emph{P} $ will be large for values of $\Lambda$
that are large enough, whereas $\Lambda$ of our Universe is
definitely small. Of course, for the early universe filled with
radiation or another ''matter'' the mentioned probability is large
nevertheless ($\emph{P}\sim 1$) but in this case we have no
inflation which is a standard solution for the flatness and
horizon problems. In the other hand, the alternative solution of
these problems can be obtained in framework of cosmologies with
varying speed of light $c(t)$ (VSL). We show that, as a matter of
principle, such quantum VSL cosmologies exist that $\emph{P}\sim
1$, $\rho_{_\Lambda}/\rho_c\sim 0.7$ ($\Lambda$-problem) and both
horizon and flatness problems are solvable without inflation.



\end{abstract}

\maketitle

\section{\label{sec:level1}Introduction}

One of the major requests concerning the quantum cosmology is a
reasonable specification of initial conditions in early universe,
that is in close vicinity of the Big Bang. There are known the
three common ways to describe quantum cosmology: the
Hartle-Hawking wave function \cite{4}, the Linde wave function
\cite{5}, and the tunneling wave function \cite{6}. In the last
case the universe can tunnel through the potential barrier to the
regime of unbounded expansion. Following Vilenkin \cite{7} lets
consider the closed ($k=+1$) universe filled with radiation
($w=1/3$) and $\Lambda$-term ($w=-1$). One of the Einstein's
equations can be written as a law of a conservation of the
(mechanical) energy: $P^2+U(a)=E$, where $P=-a\dot a$, $a(t)$ is
the scale factor, the ''energy'' $E=$ const and the potential
$$
U(a)=c^2a^2\left(1-\frac{\Lambda a^2}{3}\right),
$$
where $c$ is the speed of light; see Fig.1
\begin{figure}
\centering\leavevmode\epsfysize=5.5cm \epsfbox{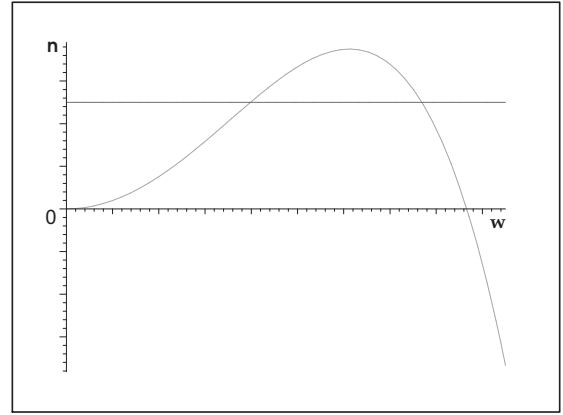}
\newline
\caption{\label{fig:epsart} Vilenkin potential with $c=$const.}
\end{figure}
The maximum of the potential $U(a)$ is located at
$a_e=\sqrt{3/2\Lambda}$ where $U(a_e)=3c^2/(4\Lambda)$. The
tunneling probability in WKB approximation can be estimated as
\begin{equation}
\emph{P}\sim\exp\left(-\frac{2c^2}{8\pi G\hbar}\int_{a'_i}^{a_i}
da\sqrt{U(a)-E}\right), \label{1}
\end{equation}
where $a'_i<a_i$ are two turning points. The universe can start
from $a=0$ singularity, expand to a maximum radius $a'_i$ and then
tunnel through the potential barrier to the regime of unbounded
expansion with the semiclassical tunneling probability (\ref{1}).
Choosing $E=0$ one gets $a'_i=0$ and $a_i=\sqrt{3/\Lambda}$. The
integral in (\ref{1}) can be calculated. The result can be written
as
\begin{equation}
 \emph{P}\sim
\exp\left(-\frac{2c^3}{8\pi G\hbar\Lambda}\right). \label{2}
\end{equation}
For the probability to be of reasonable value, for example
$\emph{P}=1/{\rm e}\sim 0.368$, one has to put $\Lambda\sim
0.3\times 10^{65}\,\, {\rm sm}^{-2}$ (see (\ref{2})). In other
words, the $\Lambda$-term must be large. However, despite this
problem, we does acquire one prise: Once nucleated, the universe
immediately begins a de Sitter inflationary expansion. Therefore
the tunneling wave function results in inflation. And the
$\Lambda$-term problem, which arises in this approach is usually
being gotten rid of via the anthropic principle.

Besides, there exists another way to estimate the (\ref{1}). If
the ''energy'' is a random variable, one can consider $\emph{P}$
as a function from both $\Lambda$ and $E$. Then, assuming $E\sim
U(a_e)=3c^2/4\Lambda$ we quickly come to conclusion that the
integral in (\ref{1}) goes to zero and $\emph{P}\to 1$. Such
universe don't experience the inflation, therefore we are unable
to classically solve the flatness and other problems. What can be
said about $\Lambda$-term? For this to answer we'll introduce the
dimensionless variable $z_0={\tilde a}_0\sqrt{\Lambda}$ where
${\tilde a}_0=10^{28}$ sm is the widely accepted modern value of
the scale factor. Assuming $H_0{\tilde a}_0=c$ ($H_0$ is the
modern value of the Hubble parameter) allows to extract from the
Einstein equation the biquadratic equation $4z_0^4-24 z_0^2+9=0,$
therefore $z_0=2.366$ (the second positive root $z'_0=0.634$ is
less then $z_e=1.225$ which approximately corresponds to the
initial value of the scale factor). Since the contribution of
$\Lambda$ to the density is $\rho_{_\Lambda}=\Lambda c^2/8\pi G$,
we can calculate the value
$\Omega_{_\Lambda}=\rho_{_\Lambda}/\rho_c$, where
$\rho_c=3H_0^2/8\pi G$ is the critical density. Upon doing this we
get $\Omega_{_\Lambda}=z_0^2/3=1.866$ instead of observed
$\Omega_{_\Lambda}=0.7$.

One can sum up all the premises as follows: (i) The semiclassical
tunneling probability for the universe to nucleate into the
inflation phase is very small for the small values of the
$\Lambda$-term; (ii) the tunneling nucleation probability is large
($\emph{P}\sim 1$) for the universe which is filled with "matter"
(radiation ad hoc) - but with the total loss of inflation. Thus we
have two different ways for the further inquiries: either to
prefer the inflation and then go on with the anthropic principle
or to find some kind of the inflation's alternates.

Among such alternatives in physics nowadays one of the most
interesting are certainly the cosmological models with varying
speed of light (VSL) \cite{1}, \cite{2} (In fact, there are many
articles about this matter. But we'd like to restrict ourselves to
consider only these ones which has been used in this work.). In
simplest case the speed of light $c =c(t)$ varies as some power of
the expansion scale factor: $c(t)=s a^n(t)$, where constant $s>0$.
Summarizing some of the promising positive features of these
models: (a) It can solve the horizon and flatness problems if
$n\le n_{_{fl}}=-(3w+1)/2$; (b) in case of
$n<n_{_\Lambda}=-3(w+1)/2$ the VSL models can solve the
$\Lambda$-problem in a early universe while inflation models can't
handle it without the aid of the anthropic principle.

Of course, these VSL models result in some shortcomings and
unusual (unphysical?) features as well \cite{3}: (1) It is not
clear how to solve the isotropy problem; (2) the quantum
wavelengths of massive particle states and the radii of primordial
black holes can grow sufficiently fast to exceed the scale of the
particle horizon; (3) the entropy problem: Entropy can decrease
with increasing time.

Keeping in mind all the above-mentioned problems we'd like,
nevertheless, to consider VSL quantum cosmology. One of
interesting observations is that the probability to find the
finite universe short after it's tunneling through the potential
barrier is $\emph{P}\sim\exp(-\beta\Lambda^{\alpha})$ with
$\alpha>0$ and $\beta>0$ for the special values of $n$ (see
below). This  means that the difference between $\emph{P}$ in VSL
and usual quantum cosmology can be very significant.

The plan of the paper looks as follows: in the next Section we'll
consider the simplest VSL model: model of
Albrecht-Magueijo-Barrow.  In Sec. III we'll study the case of
nonsingular potentials $U(a)$ (the case A, with $n>n_{_{fl}}(w)$,
see below). Although such a potentials are not fit to solve the
cosmological problems in framework of classical VSL cosmology,
they can be of interest in framework  of quantum cosmology. In
particular, as we will see, these potentials are easily result in
$\Omega_{_\Lambda}\sim 0.7$. In Sec. IV and V we'll study the
models with singular potentials: the cases B
($n_{_\Lambda}(w)<n<n_{_{fl}}(w)$) and  C ($n<n_{_\Lambda}(w)$).
We'll show that only potentials of the case B do have the ground
state and therefore do have the physical meaning. Another
interesting feature of the case B (Sec. V) is that it allows to
solve the horizon and flatness problems without the aid of
inflation. However, the $\Lambda$-problem can't be solved in this
case (on the classical level) {\it but quantum cosmology predict
$\Omega_{_\Lambda}\sim 0.7$ if ${\tilde a}_0=10^{28}$ sm with
$\emph{P}\sim 1$!} Unfortunately, in the case B we can't obtain
the universe with $w>1/9$ just after nucleation and this is
probably the major shortcoming of the model. Despite the fact that
the case C has no clear physical meaning it will be briefly
considered in Sec. V, in hope that the string theory can breathe
new life into these models (see the discussion in Sec. V). As an
another reason can be named the interesting classical behavior of
this model (see Appendix) while far from the singularity $a=0$. In
Sec. VI we'll consider two peculiar cases when $n=n_{_{fl}}$ and
$n=n_{_\Lambda}$. It will be shown that the ground state is
admitted for the first case only.

\section {Albrecht-Magueijo-Barrow VSL model}
Lets start with the Friedmann and Raychaudhuri system of equations
with $k=+1$ (we assume that $G$=const):
\begin{equation}
\begin{array}{cc}
\displaystyle{\frac{\ddot a}{a}=-\frac{4\pi
G}{3}\left(\rho+\frac{3 p}{c^2}\right)+\frac{\Lambda c^2}{3},}
\\
\\
\displaystyle{\left(\frac{\dot a}{a}\right)^2=\frac{8\pi
G\rho}{3}-\left(\frac{c}{a}\right)^2+\frac{\Lambda c^2}{3},}
\\
\\
\displaystyle{c=c_0\left(\frac{a}{a_0}\right)^n=sa^n,\qquad
p=wc^2\rho,}
\end{array}
\label{frid}
\end{equation}
where $a=a(t)$ is the expansion scale factor of the Friedmann
metric, $p$ is the fluid pressure, $\rho$ is the fluid density,
$k$ is the curvature parameter, $\Lambda$ is the cosmological
constant, $c_0$ is the modern value of the speed of light
($3\times 10^{10}$ sm/sec) and $a_0$ is the modern value of the
scale factor. Usually, this value is estimated as  $10^{28}$ sm.
However, keeping in mind that the speed of light in our model is
effectively decreasing, in fact we will choose $a_0=N\times
10^{28}$ sm with some $N>0$.

Using (\ref{frid}) we get
\begin{equation}
\dot\rho=-\frac{3\dot
a}{a}\left(\rho+\frac{p}{c^2}\right)+\frac{{\dot
c}c(3-a^2\Lambda)}{4\pi G a^2}. \label{rt}
\end{equation}
After the (\ref{rt}) solving we receive
\begin{equation}
\rho=\frac{M}{a^{3(w+1)}}+\frac{3s^2na^{2(n-1)}}{4\pi G(
2n+3w+1)}-\frac{s^2n\Lambda a^{2n}}{4\pi G(2n+3w+3)}, \label{rho}
\end{equation}
where $M$ is a constant characterizing the amount of "matter" with
given $w$. It is clear from the (\ref{rho}) that the flatness
problem can be solved in early universe  by an interval of VSL
evolution if $n < n_{_{fl}}(w)=-(1+3w)/2$, whereas the problem of
$\Lambda$-term can be solved only if
$n<n_{_\Lambda}(w)=n_{_{fl}}(w)-1=-3(w+1)/2$. The evolution
equation for the scale factor $a$ (the second equation in system
(\ref{frid})) can be written as
\begin{equation}
P^2+U(a)=E, \label{equation}
\end{equation}
where $P=-\dot a a^{-n_{_{fl}}(w)}$ is the momentum conjugate to
$a$, $E=8\pi G M/3$ and
\begin{equation}
U(a)=s^2a^{2(n-n_{_{fl}}(w))}\left(\frac{3w+1}{2(n-n_{_{fl}}(w))}-\frac{\Lambda(w+1)a^2}{2(n-n_{_\Lambda}(w))}\right).
\label{U}
\end{equation}
The expressions (\ref{rho}), (\ref{U}) are valid if $n\ne
n_{_{fl}}(w)$ and $n\ne n_{_\Lambda}(w)$. These cases will be
considered separately.

The potential $U(a)$ is the "quantum" potential from the
Wheeler-DeWitt  equation. To obtain the model with the nonzero
quantum tunneling nucleation probability we should have the
potential with the maximum. If we restrict ourselves to working
only with the positive $\Lambda$ then the case $n_{_{fl}}(w)<0$
(i.e. $w>-1/3$) will get us one maximum at
$a=a_e=\sqrt{(3w+1)/\Lambda(w+1)}$, where the function $U(a)$:
\begin{equation}
U(a_e)=\frac{s^2(3w+1)}{2(n-n_{_{fl}}(w))(n-n_{_\Lambda}(w))}\left(\frac{3w+1}{\Lambda(w+1)}\right)^{n-n_{_{fl}}(w)}.
\label{Ue}
\end{equation}
Next, as can be easily seen from the pictures, there exists three
distinguishable cases: case (A) $n>n_{_{fl}}(w)$, see Fig. 2; case
(B) $n_{_\Lambda}(w)<n<n_{_{fl}}(w)$, see Fig. 3; and case (C)
$n<n_{_\Lambda}(w)$, see Fig.4.
\begin{figure}
\centering\leavevmode\epsfysize=5.5cm \epsfbox{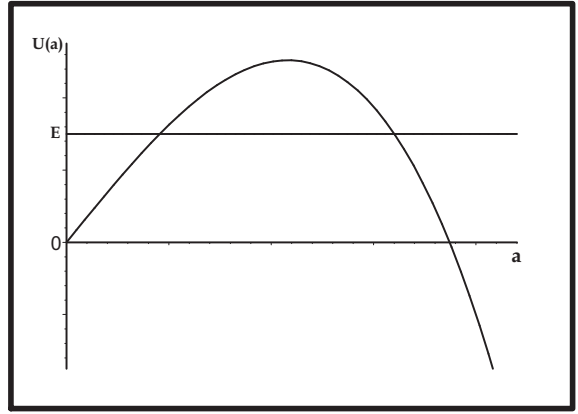}
\newline
\caption{\label{fig:epsart} The case $n>n_{_{fl}}(w)$.}
\end{figure}
\begin{figure}
\centering\leavevmode\epsfysize=5.5cm \epsfbox{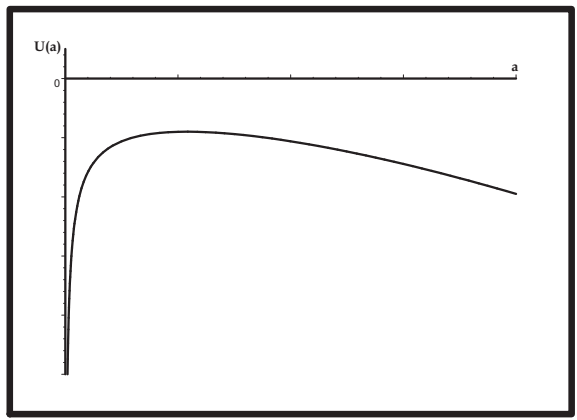}
\newline
\caption{\label{fig:epsart} The case
$n_{_\Lambda}(w)<n<n_{_{fl}}(w)$.}
\end{figure}

\section{The case A: $n>n_{_{fl}}(w)$}

This case is seemingly the one that is favorable the least. In
fact, since $n_{_\Lambda}(w)<n_{_{fl}}(w)$ then for
$n>n_{_{fl}}(w)$ we can solve neither flatness nor $\Lambda$
problems while working in the framework of the Barrow approach.
However, as we shall see, even such $n$ allows the $\Lambda$
problem to be solved - but solved in framework of quantum
cosmology.

The equation (\ref{equation}) is quite similar to equation
concerning particle of energy $E$ that is moving in potential
(\ref{U}), hence the universe in quantum cosmology can start at
$a\sim 0$, expand to a maximum radius $a'_i$ and then tunnel
through the potential barrier to the regime of unbounded expansion
with "initial" value $a=a_i$. The semiclassical tunneling
probability can be estimated as
\begin{equation}
\emph{P}\sim\exp\left(-2\int_{a'_i}^{a_i} {\mid {\tilde p(a)}\mid}
da\right), \label{P}
\end{equation}
with
$$
\begin{array}{cc}
{\mid {\tilde p(a)}\mid}=\left(8\pi
G\hbar\right)^{(n_{_{fl}}(w)-1)/2} c^{(1-3n_{_{fl}}(w))/2} \mid
P(a)\mid,\\
\\
 \mid P(a)\mid=\sqrt{U(a)-E},
\end{array}
$$
where $E\le U(a_e)$. It is convenient to write
$E=U(a_e)\sin^2\theta$, with $0<\theta<\pi/2$. After calculation
of this one we get
\begin{widetext}
\begin{equation}
\begin{array}{cc}
\displaystyle{
\emph{P}\sim\exp\left(-\frac{\sqrt{2}}{\sqrt{(n-n_{_{fl}}(w))(n-n_{_\Lambda})}}\left(8\pi
G\hbar\right)^{(n_{_{fl}}(w)-1)/2}s^{3(1-n_{_{fl}}(w))/2}\Lambda^{(3n+2)(n_{_{fl}}(w)-1)/4}W(w,\theta,n)\right)},\\
\\
\displaystyle{
W(w,\theta,n)=\int_{z'_i}^{z_i}\frac{dz}{z^{n(3n_{_{fl}}(w)-1)/2}}\sqrt{z^{2(n-n_{_{fl}}(w))}\left((n-n_{_\Lambda}(w))
(3w+1)-(n-n_{_{fl}}(w))(w+1)z^2\right)-\kappa(w,\theta,n)},}\\
\\
\displaystyle{\kappa(w,\theta,n)=\frac{(3w+1)^{(n-n_{_\Lambda}(w))}\sin^2\theta}{(w+1)^{n-n_{_{fl}}(w)}}}
 \label{ops}
\end{array}
\end{equation}
\end{widetext}
where $z=a\sqrt{\Lambda}$, $z'_i<z_i$ are the roots of expression
under the integral. As we can see, for the $-(1+3w)/2<n<-2/3$ the
probability $\emph{P}$ is increasing when $\Lambda\to 0$! In fact,
the equations (\ref{equation}) apply the restriction on the values
of $\Lambda$: the $\Lambda$ can't be too small.

To show this, lets consider the case $w=1/3$ (radiation). If $n=0$
(i.e. $c$=const) the  probability $\emph{P}$ does not vanish in
the limit of $E\to 0$, when there is no radiation and the size of
the initial universe shrinks to zero. In our situation this is not
the case: the probability $\emph{P}=0$ at $E=0$. Therefore a
newborn universe will inevitably be filled with the radiation.

In fact, for the small $E$ the (\ref{ops}) can be estimated  as
$$
\emph{P}=\exp\left(-\frac{s^3 I(n,\theta)}{\pi G\hbar
(4(n+1)^{3(n+1)/2})}\left(\frac{3(n+2)}{\Lambda}\right)^{(3n+2)/2}\right),
$$
where
$$
I(n,\theta)=\int_{x'_i(n,\theta)}^{\pi/2}dx \left((\sin
x)^{3n+1}-(\sin x)^{3(n+1)}\right),
$$
with
$$
x'_i(n,\theta)=\arcsin\left((\sin\theta)^{1/(n+1)}\sqrt{\frac{n+1}{(n+2)^{(n+2)/(n+1)}}}\right).
$$
Choosing $\theta=0.1$ and $n=-0.9$ one get
$\emph{P}=\exp\left(-0.176\times 10^{149}\sqrt{\Lambda}\right)$,
so this probability will be $\emph{P}\sim 1/{\rm e}\sim 0.37$ for
the $\Lambda\sim 0.3\times 10^{-296}$ ${\rm sm}^{-2}$. But it is
impossible due to the equation of motion (\ref{equation}).
Choosing $E\sim 0$, we get
\begin{equation}
\Lambda=\frac{3(n+2)^2}{2(n+1)}\sim 18.15\,\,{\rm sm}^{-2},
\label{18}
\end{equation}
for the $n=-0.9$, therefore $\emph{P}\sim\exp\left(-0.75\times
10^{149}\right)\sim 0$.

Thus, the probability $\emph{P}$ will be largest for $E\sim
U(a_e)$ (of course, if $E>U(a_e)$ there is no quantum tunneling at
all). So, we can choose $\theta\sim \pi/2$. In this case $z'_i\sim
z_i$ and the integral $W(w,\theta\sim\pi/2,n)\sim 0$, hence
$\emph{P}\sim 1$. If, for example, $w=1/3$ then
$$
\emph{P}\sim \exp\left(-\frac{3\sqrt{2} \epsilon^2
s^3\Lambda^{-(3n+2)/2}}{8\pi G\hbar
(n+1)(n+2)}\left(\frac{3}{2}\right)^{3n/2}\right),
$$
where $\epsilon=\pi/2-\theta\ll 1$.

Lets consider the equation (\ref{equation}) with $w=1/3$. It leads
us to equation
\begin{equation}
\begin{array}{cc}
\displaystyle{ (a{\dot
a})^2+s^2a^{2(n+1)}\left(\frac{1}{n+1}-\frac{2\Lambda
a^2}{3(n+2)}\right)=}\\
\\
\displaystyle{
\frac{s^2}{(n+1)(n+2)}\left(\frac{3}{2\Lambda}\right)^{n+1}},
\label{opppa}
\end{array}
\end{equation}
with $\sin\theta\sim 1$. Substituting $a=a_0=N c_0/H_0=N\times
10^{28}$ sm, where $H_0$ is the Hubble root we get
\begin{equation}
z_0+\frac{1}{(n+1)z_0^{n+1}}=
\left(n+2\right)\left(N^2+\frac{1}{n+1}\right), \label{z0}
\end{equation}
where $z_0=2\Lambda a_0^2/3$. The modern contribution of $\Lambda$
into the density is $\rho_{_\Lambda}=\Lambda c_0^2/(8\pi G)$. We
can define the quantities
$\Omega_{_\Lambda}=\rho_{_\Lambda}/\rho_c$, where
$\rho_c=3H_0^2/(8\pi G)$ is the critical density and
$\Omega_{_R}=\rho_{_R}/\rho_c$ where $\rho_{_R}$ is the radiation
contribution. The simple calculation results in
\begin{equation}
\Omega_{_\Lambda}=\frac{z_0}{2 N^2},\qquad
\Omega_{_R}=\frac{z_0^{-(n+1)}}{(n+1)(n+2)N^2}.
\label{Om2}
\end{equation}
Note, that in the beginning $z_i\sim 1$. Now we can solve
(\ref{z0}) for any given $n$ (from the interval $-1<n<-2/3$) and
$N$ in order to find $z_0$ which should be next substituted into
the (\ref{Om2}). The explicit results are presented in Table 1.
\begin{table}
\caption{\label{tab:table2}The table of values of $z_0$,
$\Omega_{_\Lambda}$ and $\Omega_{_R}$ for some of the $n$ from the
interval $(-1;-2/3)$ and a few values of $N$. The most of values
of $\Omega_{_\Lambda}$ are adjacent to the range $0.7$, and that
is quite consistent with the observational data. }
\begin{ruledtabular}
\begin{tabular}{ccccc}
n & N &$z_0$ &$\Omega_{_\Lambda}$&$\Omega_{_R}$\\
\hline -0.9&1& 3.2 & 1.6 & 8.09\\
-0.9&2& 7.19 & 0.899 & 1.87\\
-0.9&3& 13.7 & 0.732 & 0.78\\
-0.9&4& 21.23 & 0.664 & 0.42\\
-0.9&5& 31.42 & 0.628 & 0.25\\
-0.9&10& 114.78 & 0.574 & 0.06\\
-0.9&100& 1.1$\times 10^{4}$ & 0.55 & 0.36$\times 10^{-3}$\\
\hline -0.8 & 1 & 3.25 & 1.63 & 3.29\\
-0.8 & 2 & 7.45 & 0.93 & 0.7 \\
-0.8 & 3 & 13.84 & 0.77 & 0.27 \\
-0.8 & 4 & 22.52 & 0.7 & 0.14 \\
-0.8 & 5 & 33.52 & 0.67 & 0.08 \\
-0.8 & 10 & 124.09 & 0.62 & 0.02 \\
-0.8 & 100 & 1.2$\times 10^{4}$ & 0.6 & 6$\times 10^{-5}$ \\
\hline -0.7 & 1 & 3.3 & 1.65 & 1.79 \\
-0.7 & 2 & 7.73 & 0.97 & 0.35 \\
-0.7 & 3 & 14.54 & 0.81 & 0.13 \\
-0.7 & 4 & 23.85 & 0.75 & 0.06 \\
-0.7 & 5 & 35.69 & 0.71 & 0.04 \\
-0.7 & 10 & 133.57 & 0.67 & 6$\times 10^{-3}$ \\
-0.7 & 100 & 1.3$\times 10^{4}$ & 0.65 & 6$\times 10^{-5}$ \\
\end{tabular}
\end{ruledtabular}
\end{table}
For all the cases the value $\Lambda\sim 10^{-55}$ ${\rm
sm}^{-2}$, which is much less then (\ref{18}).

As we can see, the values of $\Omega_{_\Lambda}$ lies in most near
the range $0.7$. And this is in quite good consent with the
observational data.

\begin{figure}
\centering\leavevmode\epsfysize=5.5cm \epsfbox{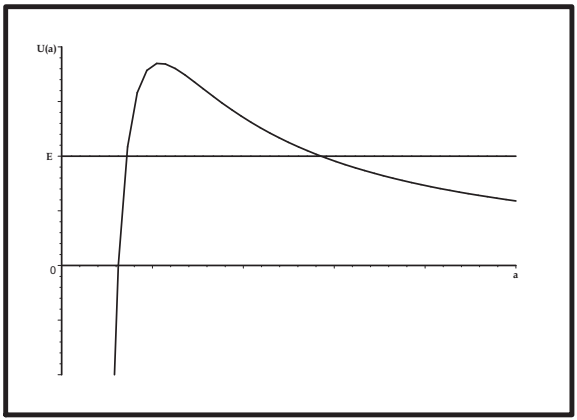}
\newline
\caption{\label{fig:epsart} The case $n<n_{_\Lambda}(w)$.}
\end{figure}

\section{An existence of ground states for singular potentials (The case B)}

It is interesting to ask: what can be said about the value of
$U(a)$ when $a=0$? It is easy to see from (\ref{U}), that
(depending on the values of $w$ and $n<0$), the potential $U(a)$
can take on one of two different values: either $0$ (Fig. 2) or
$-\infty$, (Fig. 3 and Fig. 4). Since we consider only $w>-1/3$
values, the second case is valid for every $n$ satisfying the
inequality $n<n_{_{fl}}(w)<0$. Thus, it leads us to another
question: what can be the possible meaning of the potential which
at $a\to 0$ is unbounded from below? It seems that such universe
is able to just roll down towards small values of $a$ (where the
potential is tending to minus infinity) instead of any tunneling
to large values.

This situation can in fact be alleviated if the considered
potential $U(a)$ has the ground state. Indeed, one can imagine the
fictitious particle with some energy and coordinate $a(t)$ in the
potentials (\ref{U}) (see Fig.3 and Fig. 4) rolling down towards
small values of $a$. The main problem is: whether the quantum
mechanical energy spectrum of $U(A)$ is unbounded below? If not,
then it does admits the ground state and hence can have the
physical meaning.

To find such a potential lets suppose that our fictitious particle
is located in a small region $a$ near the singularity $a=0$, with
the momentum $P$. One can use the Heisenberg uncertainty relation
as
\begin{equation}
P\,a\sim \left(8\pi G\hbar\right)^{(1-n_{_{fl}}(w))/2}
c^{(3n_{_{fl}}(w)-1)/2}.
\label{Heisen}
\end{equation}
Using (\ref{Heisen}), and (\ref{frid}) (or (\ref{equation})) one
get for the $a\to 0$
$$
\begin{array}{cc}
\displaystyle{ E=P^2+U(a)\to}\\
\\
\displaystyle{\to \frac{Z^2}{a^{2-n(3n_{_{fl}}(w)-1)}}+\frac{s^2
n_{_{fl}}(w)}{(n_{_{fl}}(w)-n)a^{2(n_{_{fl}}(w)-n)}}},
\end{array}
$$
where $Z^2=\left(8\pi G\hbar\right)^{1-n_{_{fl}}}
s^{3n_{_{fl}}-1}$, and
\begin{equation} n<n_{_{fl}}(w)<0.
\label{nerv0}
\end{equation}
Therefore the energy spectrum will be bounded below if
\begin{equation}
\left(3n+2\right)\left(n_{_{fl}}(w)-1\right)<0. \label{nervy}
\end{equation}
and (\ref{nerv0}) are valid. This situation is represented
graphically on the Fig. 5.
\begin{figure}
\centering\leavevmode\epsfysize=5.5cm \epsfbox{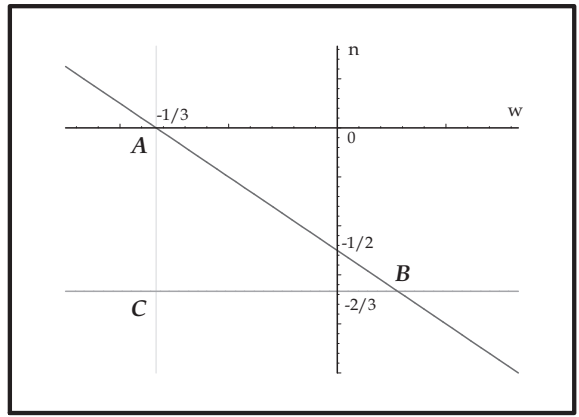}
\newline
\caption{\label{fig:epsart} The ground state exists for $w$ and
$n$ from the interior of the triangle ABC.}
\end{figure}
It is easy to see that the conditions of ground state existence
can be satisfied by case B only (see Fig. 3). It is also important
to note that the left side of the (\ref{nervy}) (accurate to the
coefficient $1/4$) is actually the power of $\Lambda$ in the
probability (\ref{ops})${}^{[1]}$~\footnotetext[1]{The power of
$\Lambda$ is similar for all the values of $n$ except for the
cases $n=n_{_{fl}}(w)$ and $n=n_{_\Lambda}(w)$.}. But doesn't it
means that those sensible potentials which are unbounded from
below at $a\to 0$ all result in semiclassical tunneling nucleation
probability, strongly suppressed for small values of $\Lambda$,
just as in the models with $c$=const? In reality, the through
examination of the model shows that the situation is much better
than it first looks. As we shall see, the universes which have
nucleated with the probability $\emph{P}\sim 1$ would have the
$\Omega_{_\Lambda}\sim 0.7$ for the $N\sim 1$!

To show this, lets choose $w=0$ (unfortunately, the case $w=1/3$
is inaccessible for the case B, since the maximum value is
$w=1/9$, see Fig. 5) and $E=U(a_e)\cosh\theta$. It is obvious that
$\emph{P}\sim 1$ for sufficiently small $\theta$. Choosing
$\theta\to 0$ one can estimate the probability as
$$
\emph{P}\sim
\exp\left(\frac{4s^{9/4}\theta^2}{(2n+1)(2n+3)\left[8\pi
G\hbar\Lambda^{(3n+2)/2}\right]^{3/4}}\right).
$$
Now, lets consider the equation (\ref{equation}) with $w=0$. We
get
\begin{equation}
N^2+\frac{1}{2n+1}=\frac{1}{2n+1}\left(z_0^2+\frac{2}{(2n+3)z_0^{2n+1}}\right),
\label{z1}
\end{equation}
where
$$
z_0=\frac{Nc_0\sqrt{\Lambda}}{H_0}=a_0\sqrt{\Lambda}.
$$
The equation (\ref{z1}) is the analog of the (\ref{z0}) for the
case $w=0$. Finally, $\Omega_{_\Lambda}=z_0^2/(3N^2)$. It is
necessary to remember that for case $w=0$ we have
$$
-\frac{2}{3}<n<-\frac{1}{2}.
$$
Since $2n+1<0$ the value $N$ is bounded above $N<N_{max}(n)$ (see
(\ref{z1})). For example $N_{max}(-0.65)=1.826$,
$N_{max}(-0.6)=2.236$, $N_{max}(-0.55)=3.162$. Solving (\ref{z1})
for given $n$ and $N$ one get $z_0$ and $\Omega_{_\Lambda}$ (see
Tabl. 2)
\begin{table}
\caption{\label{tab:table2}A table of values of $z_0$ and
$\Omega_{_\Lambda}$ for some $n$ from the interval $(-2/3;-1/2)$
and a few values of $N$.}
\begin{ruledtabular}
\begin{tabular}{cccc}
n & N &$z_0$ &$\Omega_{_\Lambda}$\\
\hline -0.65& 1 & 1.416 & 0.669\\
-0.65 & 1.4 & 1.298 & 0.286 \\
-0.65 & 1.8 & 1.115 & 0.128 \\
\hline -0.6 & 1 & 1.411 & 0.663 \\
-0.6 & 1.5 & 1.313 & 0.255 \\
-0.6 & 2 & 1.16 & 0.112 \\
\hline -0.55 & 1 & 1.411 & 0.663 \\
-0.55 & 2 & 1.296 & 0.14 \\
-0.55 & 3 & 1.077 & 0.043 \\
\end{tabular}
\end{ruledtabular}
\end{table}

It is really astonishing that case $N=1$ (i.e. $a_0=10^{28}$ sm)
results in $\Omega_{_\Lambda}\sim 0.7$, this result being
strikingly consent with the observational data. Besides, one can
choose to put $\Omega_{_\Lambda}=0.7$ instead and then solve
(\ref{z1}) to find $N(n)$. As a result one will get
$N(-0.65)=0.98$, $N(-0.6)=0.976$, $N(-0.55)=0.975$. Thus, in
contrast to the case A, the universe bearing a maximum probability
of nucleation via quantum tunneling and with the modern value of
scale factor near $10^{28}$ sm must has $\Omega_{_\Lambda}\sim
0.7$.

In conclusion to this section, we should discuss one point of the
model, that can be somehow of disturbance for us. The problem that
is at issue arises from the fact, that $\emph{P}\sim 1$ if
$\theta\sim 0$, i.e. $E\sim U(a_e)<0$. This means that the
constant characterizing the amount of "matter" with $w=0$ (''the
mass of dust'') is negative. We have full right to ask whether
such statement is physically consistent. To be more exact, one can
be afraid of facing the violation of the weak energy condition
$\rho>0$, $\rho+p/c^2>0$. Fortunately, in case of the considered
example such misgivings tends to be groundless. Using (\ref{rt})
we can get
$$
\rho_0=\frac{c_0^2\Lambda\left(3z_0^{-2n}+6n^2z_0+9nz_0-n(2n+1)z_0^3\right)}{4\pi
G(2n+1)(2n+3)z_0^3}, \label{rho0}
$$
where $\rho_0$ is a modern value of total density. Using the Tabl.
2 one can verify that for those values of $n$ and $w$ that are of
interest for us $\rho_0>0$.
\section{The case C}
As we have shown, the conditions of the ground state existence for
singular at $a\to 0$ potentials can only be satisfied for
potentials from the case B. With this in account, we can come to a
very uncomfortable conclusion that the case C (and a rest of case
B) has no any physical meaning.

However, in order to make these potentials physical, there still
exists but one loophole. This loophole follows from a strings
theory prediction that states the existence of $a_{min}$ -- the
minimal spatial scale. Due to the strings theory, there is no
sense in considering the physics at $a<a_{min}$. If this is true,
then we should not deny the possibility of potentials $U(a)$ with
$n<n_{_{fl}}(w)<0$ to be the physical potentials with existing
ground state - this, of course, being just a speculation. Keeping
this in mind, let us now consider the model with $w=1/3$,
$n_{_{fl}}=-1$, $n=-1-m$, and $m>0$.

In this case the semiclassical tunneling probability has a form
$\emph{P}\sim \exp(-S)$ with
$$
S=\frac{s^3\Lambda^{(3m+4)/2}}{4\pi G\hbar
3^{(m+1)/2}\sqrt{m(m+1)}}\int_{z'_i}^{z_i}
\frac{dz\sqrt{F_m(z,\theta)}}{z^{3m+5}},
$$
where
$$
F_m(z,\theta)=-2^{m+1}\sin^2\theta\, z^{2(m+1)}+2\times 3^m (m+1)
z^2-m 3^{m+1},
$$
$z$ is a dimensionless quantity and $z'_i$, $z_i$ are the turning
points, i.e. two real positive solutions of the equation
$F_m(z,\theta)=0$ for given $\theta$ (it is easy to see that the
equation $F_m(z,\theta)=0$ does have two such solutions at
$0<\theta<\pi/2$).

If $m$ is the whole number then the expression for the $\emph{P}$
has more simple form. For example
$$
\emph{P}_1\sim\exp\left(-\frac{s^3 \Lambda^{7/2}\sin\theta }{6\pi
G\hbar\sqrt{2}}\int_{z'_i}^{z_i}\frac{dz}{z^8}\sqrt{(z^2-{z'_i}^2)(z_i^2-z^2)}\right),
$$
with
$$
z'_i=\frac{\sqrt{3}}{2\cos(\theta/2)},\qquad
z'_i=\frac{\sqrt{3}}{2\sin(\theta/2)}.
$$
\newline
\newline
This expression can be calculated exactly:
\begin{widetext}
$$
\emph{P}_1\sim\exp\left(-\frac{s^3\Lambda^{7/2}\sin\theta
J(\theta)}{6\sqrt{2}\pi G\hbar}\right),\qquad
J(\theta)=\frac{1}{105}\left(\frac{2\sin(\theta/2)}{\sqrt{3}}\right)^5
\left[\Delta(\theta)
\Pi\left(\mu^2;\frac{\pi}{2}{\backslash}\arcsin\mu\right)-
2\left(2\lambda^4-\lambda^2+2\right) \textrm{K}(\mu^2)\right],
$$
\end{widetext}
with $\mu^2=\cos\theta/\cos^2(\theta/2)$,
$\lambda=\cot(\theta/2)$, $\Delta(\theta)=
(8\lambda^4-13\lambda^2+8)/\cos^2(\theta/2)$,
 $\Pi$ and $\textrm{K}$ are complete
elliptic integrals of the first  and third  kinds correspondingly
(see \cite{8})

Similarly, $\emph{P}_2\sim\exp(-S_2)$, with
$$
S=\frac{s^3 \Lambda^5\sin\theta }{18\pi
G\hbar}\int_{z'_i}^{z_i}\frac{dz}{z^{11}}\sqrt{(z^2+z_1^2)(z^2-{z'_i}^2)(z_i^2-z^2)},
$$
where
$$
\begin{array}{cc}
\displaystyle{
z_1=\sqrt{\frac{3}{\sin\theta}\cos\left(\frac{\theta}{3}-\frac{\pi}{6}\right)},\qquad
z'_i=\sqrt{\frac{3}{\sin\theta}\sin\frac{\theta}{3}},}\\
\\
\displaystyle{
z_i=\sqrt{\frac{3}{\sin\theta}\cos\left(\frac{\theta}{3}+\frac{\pi}{6}\right)}},
\end{array}
$$
and so on.

Since the case C is still questionable, we will restrict ourselves
to the examples above. Note, however, that such models can be
quite interesting in classical (non-quantum) cosmology (see
Appendix). Another example of such calculations can be found in
\cite{Yurov}.

\section{Peculiar cases}

If $n=-(3w+1)/2$ then
\begin{equation}
U(a)=s^2\left(1+(3w+1)\log(\lambda
\sqrt{\Lambda}a)-\frac{\Lambda(1+w)a^2}{2}\right), \label{Uh1}
\end{equation}
with $P=-{\dot{a}}/a^n$, and some $\lambda$, that is
dimensionless.

Similarly, if $n=-3(w+1)/2$ the potential $U(a)$ takes form:
\begin{equation}
U(a)=-s^2\left(\frac{3w+1}{2a^2}+\Lambda\left(\frac{1}{3}+(w+1)\log(\lambda
\sqrt{\Lambda}a)\right)\right), \label{Uh2}
\end{equation}
with $P=-{\dot{a}}/a^{n+1}$. Now we should study these potentials
concerning the existence of the ground state. The calculations
result in following conclusions: The potential (\ref{Uh1}) does
have a ground state for the $-1/3<w<1/9$ whereas the potential
(\ref{Uh2}) is missing it wholly. Thus, we'll further consider
only the case (\ref{Uh1}). This potential has one maximum at
$a_e=\sqrt{(3w+1)/(1+w)\Lambda}$ and
$$
U(a_e)=\frac{s^2}{2}\left(1-3w+(1+3w)\log\frac{\lambda^2(1+3w)}{1+w}\right).
$$
Substituting  $E\sim U(a_e)$ (in order to obtain $\emph{P}\sim 1$)
into the (\ref{equation}) we get the following equation
\begin{equation}
(1+w)z_0^2-(1+3w)\log\frac{z_0^2}{z_e^2}=2N^2+1+3w, \label{privet}
\end{equation}
where $z_0=\sqrt{\Lambda}N\times 10^{28}$,
$z_e=\sqrt{(3w+1)/(w+1)}$. The value of $a_e=z_e/\sqrt{\Lambda}$
is the initial value of scale factor (after the tunneling) to high
precisions. Now we can solve (\ref{privet}) for given $w$ and $N$.
The results of such a calculation are presented in Tabl. 3.
\begin{table}
\caption{\label{tab:table2}A table of values of $z_0/z_e$ and
$\Omega_{_\Lambda}$ for some $w$ from the interval $(-1/3;1/9)$
and a few values of $N$. All calculations are done for potential
(\ref{Uh1}).}
\begin{ruledtabular}
\begin{tabular}{cccc}
w & N &$z_0/z_e$ &$\Omega_{_\Lambda}$\\
\hline -0.3& 1 & 4.918 & 1.152\\
-0.3 & 5 & 22.52 & 0.966 \\
-0.3 & 10 & 44.89 & 0.957 \\
-0.3 & 100 & 447.2 & 0.952 \\
\hline -0.2 & 1 & 2.844 & 1.349 \\
-0.2 & 5 & 11.44 & 0.872 \\
-0.2 & 10 & 22.52 & 0.845 \\
-0.2 & 100 & 223.6 & 0.834 \\
\hline -0.1 & 1 & 2.361 & 1.446 \\
-0.1 & 5 & 8.762 & 0.796 \\
-0.1 & 10 & 17.1 & 0.758 \\
-0.1 & 100 & 169 & 0.741 \\
\hline 0 & 1 & 2.123 & 1.502 \\
0 & 5 & 7.417 & 0.733 \\
0 & 10 & 14.36 & 0.688 \\
0 & 100 & 141.5 & 0.677 \\
\hline 0.1 & 1 & 1.98 & 1.536 \\
0.1 & 5 & 6.57 & 0.681 \\
0.1 & 10 & 12.7 & 0.63 \\
0.1 & 100 & 124.1 & 0.607 \\
\end{tabular}
\end{ruledtabular}
\end{table}
We can see that $\Omega_{_\Lambda}\sim 0.7$ when $w=-0.1$,
$N>100$; $w=0$, $N\sim 5-10$; $w=0.1$, $N\sim 3-5$.

\section{Conclusion}
As we have seen, the semiclassical tunneling nucleation
probability in the VSL quantum cosmology is really different from
the one in the quantum cosmology with $c$=const. The most
interesting distinction lies in capability of the VSL model to
provide via the quantum nucleation the flat universe with
$\emph{P}\sim 1$ and the horizon problem solved. Moreover, the VSL
model here is the only tool of obtaining the solution of both the
flatness and horizon problems without the aid of inflation, since
it is not quite clear how to obtain the inflation in universe
where the ''matter'' energy density is greatly exceeding those of
the vacuum. And as the additional prize of the model we get
$\Omega_{_\Lambda}\sim 0.7$ without much of an effort.

However, it would be too prematurely to say that VSL quantum
cosmology is indeed the actual panacea for the $\Lambda$-mystery
and another cosmological problems as well. First of all, as we
have seen, the case B, that is the most promising one, fails to
describe quantum tunneling into the classical state with $w=1/3$.
The validity of the WKB wave function in general is the model's
second problem. And also there are the omitted pre-exponential
factors which can be nevertheless essential for the analysis of
the vicinities of the turning points.

But, even with this in account, the shown difference of $\emph{P}$
in the VSL and the usual quantum cosmology seems very interesting
and also very significant.

\begin{acknowledgments}
After finishing this work, we learned that T.Harko, H.Q.Lu,
M.K.Mak and K.S.Cheng \cite{Harko}, have independently considered
the VSL tunneling probability in both Vilenkin and Hartle-Hawking
approaches.  The interesting conclusion of their work is that at
zero scale factor the classical singularity is no longer isolated
from the Universe by the quantum potential but instead classical
evolution can start from arbitrarily small size. In contrast to
\cite{Harko}, we attract attention to the problem of
$\Lambda$-term and the possibility to obtain the flat universe
without horizon problem but filled with ''matter'' for which
$\emph{P}\sim 1$.

We'd like to thank Professor Harko for useful information about
the article \cite{Harko}. Research has been partially supported by
"Integration"  Grant N$\Phi$ 0032/1242.
\end{acknowledgments}

\appendix

\section{}

Lets take $w=1/3$, $n=-2-m$, for the $m\ge 0$. Surely, in this
case we have no ground state. In spite of this sad circumstance
we'll still consider this model far from the $a=0$ where this
model looks as a wholly satisfactory one.

Substitution of the (\ref{rho}) into the first equation of system
(\ref{frid}) yields

\begin{equation}
{\ddot
a}=\frac{1}{a^3}\left[-E+\frac{s^2}{a^{2m}}\left(-\frac{m+2}{(m+1)a^2}+\frac{2(m+1)\Lambda}{3m}\right)\right].
\label{A1}
\end{equation}

Thus we have the following situations:
\newline
1. If
$$
0<a^2\ll \frac{3m(m+2)}{2\Lambda(m+1)^2},
$$
then the curvature term is the dominating one and ${\ddot a}<0$.
\newline
2. If
\begin{equation}
 \frac{3m(m+2)}{2\Lambda(m+1)^2}\ll a^2\ll {\tilde
a}^2\equiv \left(\frac{2s^2(m+1)\Lambda}{3mE}\right)^{1/m},
\label{A2}
\end{equation}
then the dominating term is $\Lambda$-term and $\ddot a>0$ during
this time.
\newline
3. If
$$
a^2\gg \left(\frac{2s^2(m+1)\Lambda}{3mE}\right)^{1/m},
$$
then the radiation term is the dominating one and ${\ddot a}<0$.

There are two way to interpret the region (A2). The first way is
to conclude that we have cosmological inflation in early universe.
This is possible when $0<m\ll 1$. In this case we can evaluate the
number of e-foldings  $\Delta N$ during the region (A2) as
\begin{equation}
 \log
m\sim -2 m\Delta N,\qquad  \Lambda\gg \frac{3Em}{2s^2}\sim
\frac{3Em}{2c_0^2a_0^4}. \label{A3}
\end{equation}
 If $\Delta N\sim 60$ then
$m\sim 0.029$; if $\Delta N\sim 100$ then $m\sim 0.0197$. To
evaluate $E$ one can use the well-known expression for the
Friedmann integrals \cite{Cher},
$$
A(w)=\left[\left(\frac{1+3w}{2}\right)^2 E\right]^{1/(1+3w)}.
$$
Since $A(1/3)=3\times 10^{36}$ ${\rm sm}^2/{\rm sec}$, we get
$E=0.9\times 10^{73}$ ${\rm sm}^4/{{\rm sec}^2}$. The substitution
of $A(1/3)$ into the (\ref{A3}) results in
$$
\Lambda\gg 0.435\times 10^{-61}\,\,{\rm sm}^{-2},\qquad \Lambda\gg
0.296\times 10^{-61}\,\,{\rm sm}^{-2},
$$
for the $\Delta N=60$ and $\Delta N=100$.

But do we really need inflation in the VSL models? The question is
not quite clear. On the one hand, VSL models can solve fundamental
cosmological problems (horizon and  flatness problems) without
inflation - and what is more, these models can solve
$\Lambda$-problem whereas inflations can't do it without the
anthropic principle. On the other hand, the simplest case of VSL
cosmological models, which is the subject of this article, is
facing with the isotropy problem \cite{3}. But, as we have seen,
VSL model results in inflation with exit naturally so it will be
incorrect to oppose VSL models and the inflation.

Another way to interpret the region (\ref{A2}) is to identify this
region with the modern acceleration of universe. This is possible
if $m$ is sufficiently large. Let us make a crude guess. According
to modern observations we can write ${\ddot a_0}/a_0=5.6\pi
G\rho_c/3$ where $\rho_c=10^{-29}$ ${\rm {gramme/sm^{3}}}$. If the
modern value of $a_0\sim{\tilde a}$ (see the inequality
(\ref{A2})) then
\begin{equation}
 \Lambda=\frac{3mE}{3c_0^2(m+1)a^4_0}.
 \label{A4}
\end{equation}

 From the (\ref{A1}) we have
$$
\ddot a_0\sim\frac{2(m+1)\Lambda s^2}{3ma^{2m+3}_0}
$$
if the $\Lambda$-term is dominating one.  Substituting (\ref{A4})
gets us
$$
\Omega_{_{\Lambda}}=\frac{\Lambda c_0^2}{8\pi G\rho_c}\sim
\frac{0.35 m}{m+1},\qquad a_0\sim{}^4\sqrt{\frac{3E}{5.6\pi G
\rho_c}}.
$$
Thus, if $m\gg 1$ then $\Omega_{_{\Lambda}}\sim 0.35$,
$\Lambda\sim 0.2\times 10^{-56}$ ${\rm sm}^{-2}$, $a_0\sim
10^{27}$ sm. And these values are seems to be quite reasonable.

$$
{}
$$
\bibliography{apssamp}

\vfill \eject

\end{document}